# An FPGA Implementation of Convolutional Spiking Neural Networks for Radioisotope Identification

Xiaoyu Huang[1], Edward Jones[2], Siru Zhang[3], Shouyu Xie[1], Steve Furber[2], *Fellow, IEEE*, Yannis Goulermas[3], Edward Marsden[4], Ian Baistow[4], Srinjoy Mitra[1], Alister Hamilton[1]
[1]University of Edinburgh, UK, [2]University of Manchester, UK, [3]University of Liverpool, UK, [4]Kromek Group plc, UK
E-mail: xiaoyu.huang@ed.ac.uk, alister.hamilton@ed.ac.uk

*Abstract*—This paper details FPGA implementation methodology for Convolutional Spiking Neural Networks (CSNN) and applies this methodology to low-power radioisotope identification using high resolution data. A power consumption of 75 *mW* has been achieved on an FPGA implementation of a CSNN, with the inference accuracy of 90.62% on a synthetic dataset. The chip validation method is presented. Prototyping was accelerated by evaluating SNN parameters using SpiNNaker neuromorphic platform.

## I. Introduction

Radioisotope detection and identification (ID) play an important role in national security and are key to countering the threat of nuclear weapons and dirty bombs. Conventional radioisotope ID methods are mostly frame-based, meaning that gamma photon detection events with different energies are counted to generate an energy histogram over a predefined integration time. Each target radioisotope has a characteristic energy histogram meaning that radioisotope ID can be carried out as a histogram classification task. Many algorithms have been studied and applied to this task [1][2][3].

For dynamic radioisotope ID a key drawback of the frame-based method is high constant power consumption. No matter whether there is target radioisotope, the data has to be continuously integrated to generate the histogram and processed by algorithms with computationally expensive arithmetic operations such as floating-point Multiplication-Accumulates (MACs). Furthermore, those algorithms are normally implemented on power-hungry computing architecture such as CPU or GPU, which are not always the most suitable for power constrained edge devices. In [4][5], this unnecessary power overhead of the frame-based methods for radioisotope ID was pointed out and an event-based Spiking Neural Network (SNN) architecture was proposed and implemented on a Field-Programmable Gate Array (FPGA). In an event-based manner, the always-on energy histogram generation process is removed and processing only takes place when detection events happen.

In [5] a proof-of-principle was presented, in which a shallow fully-connected SNN architecture demonstrated the high power efficiency and high accuracy on the radioisotope ID task by using an event-based SNN. However, the input to this model was resampled to only cover 100 energy channels in the input layer, limiting the system's application in cases where isotopes have similar histograms or where input has a low signal-to-noise ratio. This limitation is due to the trade-off between data dimension, power and area overhead. In the fully-connected structure, where the architecture is such that all the neurons in one layer are connected to the neurons in the next layer, the area and power overhead increases heavily for higher dimensional data.

Based on the fact that the main peak energy channels of each target radioisotope are often densely clustered around particular energy channels, a Convolutional Spiking Neural Network (CSNN) could be a better candidate for high dimensional data. In this kind of network the naïve resampling of the input energy channels is replaced by learnt filters that perform feature extraction while sharing weights. This keeps both memory and computational demands lower than in a fully-connected implementation.

In this paper, an FPGA implementation methodology for a CSNN architecture for low-power and high resolution radioisotope identification is presented in detail. Sections II-V presents the CSNN architecture and the methodology of the hardware implementation. Section VI describes the experiments undertaken to evaluate this implementation. Finally Section VII discusses the findings and outlines future work.

## II. Convolutional Spiking Neural Networks

### A. CSNN Architecture

As shown in the architecture illustration in Fig. 1, the CSNN supports 1,024 energy channels. In the convolutional layer, there are four filters with a filter size of five. It has a pooling layer with a size of 16 and a fully-connected output layer with eight neurons for each radioisotope label.

Table I shows the comparison between this convolutional structure and the fully-connected structure in [5]. The CSNN in this work supports more than ten times higher data dimensions with less than half of the weights.

TABLE I.    Comparisons

|                      | [5]             | This work     |
|----------------------|-----------------|---------------|
| Structure            | Fully-connected | Convolutional |
| Input Neuron Number  | 100             | 1,024         |
| Total Neuron Number  | 148             | 5,368         |
| Weight Number        | 4,320           | 2,036         |
| Operations Per Event | Max. 360        | Max. 160      |

### B. CSNN Training and Conversion

The process of training and conversion was made up of three steps: training, quantisation and network conversion.

In the training step the traditional Convolutional Artificial Neural Network (CANN) parameters were fine-tuned using quantisation aware training available in the TensorFlow Model Optimization Toolkit via backpropagation algorithm on the histogram dataset. In the quantisation step the weights were quantised to 8-bit signed integers.

In the network conversion step, the principle is to approximate the activation value of the neuron in CANN by the firing rate of the same neuron in CSNN [6]. This approach is particularly applicable here as process of radioactive decay can itself be modelled as Poisson process.

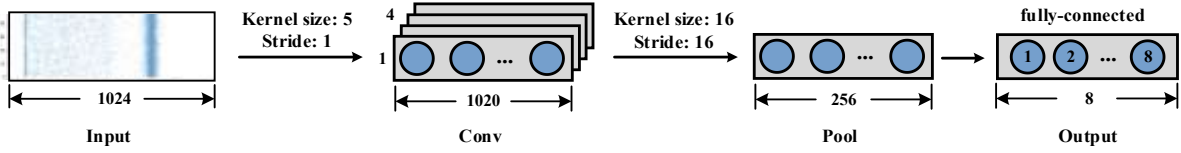

Fig. 1: An illustration of CSNN architecture.

## III. CSNN Implementation Microarchitecture

### A. Convolutional Layer

The convolutional layer of the CSNN is implemented in the microarchitecture shown in Fig. 2. The Address Event Representation (AER) protocol [7] is for communication between each layer, a ROM memory is used for the storage of all the synaptic weights, and a block RAM (BRAM) is used for the storage of the membrane voltages of the neurons.

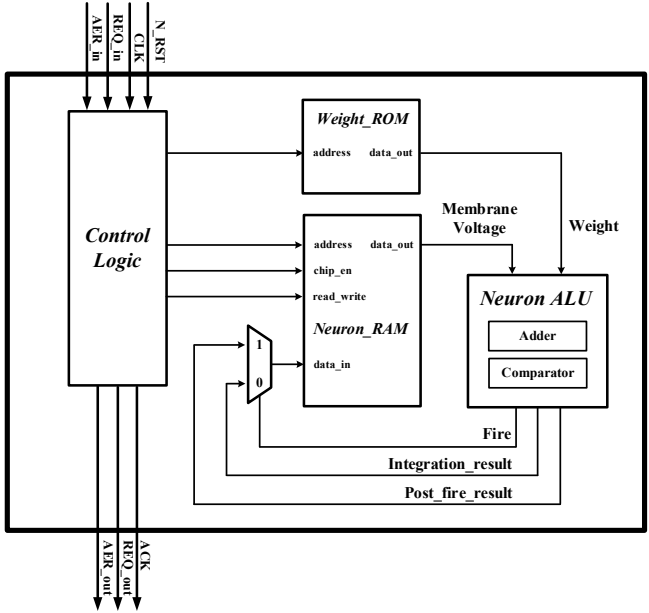

Fig. 2 Microarchitecture of the convolutional layer

Each neuron in the convolutional layer is triggered by the same pre-synaptic neuron and processed one-by-one in a Time Division Multiplexing (TDM) manner. In this architecture, there are 20 neurons connected to the same pre-synaptic neuron in the input layer, this comes from there being 4 filters in the convolutional layer, with each convolutional filter being of size 5. The sequence of operations and data flow is managed by the state machine in the Control Logic block. Table I shows the worst case maximum operations per each input spike event. Compared with the fully-connected structure in [5], the convolutional structure saves half of the operations, which leads to potential dynamic power savings.

Fig. 3 shows the timing diagram for an example of one filter operation behaviour in the convolutional layer. Same process repeats three times for another three filters in series in a TDM manner.

### B. Pooling Layer and Fully-connected Layer

The pooling layer is used to reduce the dimensionality before the fully-connected output layer. Average pooling was chosen because of the ease with which it can be implemented in a rate-based spiking model [6] and its comparable performance with "max" pooling. The fully-connected output layer is implemented also in TDM manner based on the microarchitecture in [5].

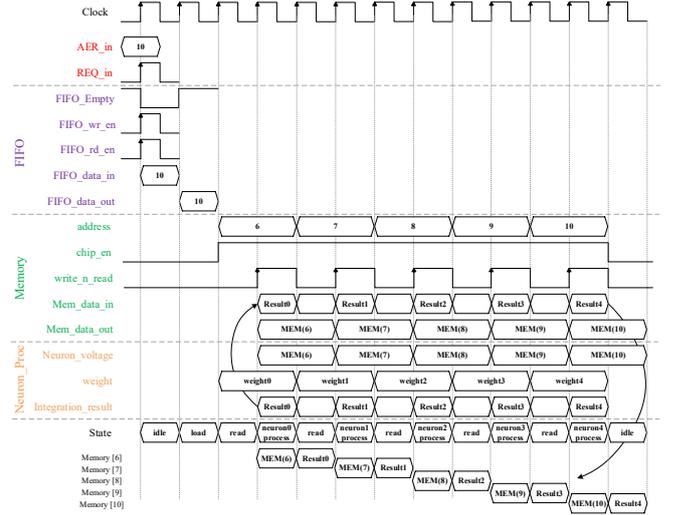

Fig. 3 Timing diagram of one transaction example in hidden layer

## IV. Neuron Model

The neuron model used in this work is based on a simplified Integrate-and-Fire (IF) model. SNNs with the similar simplifications in [5][8] demonstrated high inference performance on Radioisotope and MNIST dataset.

$$V(t) = V(t-1) + \sum_i x_i(t-1)w_i + L, \quad (1)$$

$$V(t) = V(t-1) + w_i, \quad (2)$$

$$if\ V(t) \geq V_{thr}, \quad Fire\ and\ Reset\ V(t) = V(t) - V_{thr}, \quad (3)$$

$$if\ V(t) < V_{min}, \quad Reset\ V(t) = V_{min} \quad (4)$$

The dynamics of the IF model are described by equations (1-4) [8]. The current membrane voltage $V(t)$ of the neuron equals to the sum of three components. They are the previous membrane voltage, sum of all the active post-synaptic weight contribution and the leak voltage. In SNN, the $x_i(t-1)$ represents either the presence (1) or absence (0) of a spike from the pre-synaptic neuron, and the leak parameter $L$ is zero in this work.

In this application, since the photon detector can only capture a single gamma photon event within each time step and all the neuron computations are processed in TDM manner, each membrane voltage update can be simply implemented by adding the weight of connection with the pre-synaptic neuron that fired, this can be described by equation (2).

The reset by subtraction approach is applied when the membrane voltage is above the threshold, which is shown in equation (3). Compared with the reset by zero manner, it is more suitable for ANN-to-SNN conversion, which is reported by [6] and also proved by our experiments.

To avoid the underflow and wraparound of the membrane voltage register, the minimum boundary voltage is set and the membrane voltage is reset to the minimum $V_{min}$ if it is under the boundary.

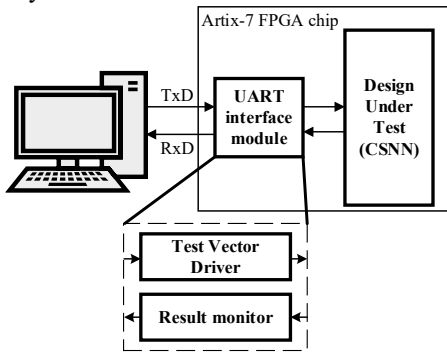

Fig. 4 Chip validation structure

## V. VALIDATION

Once the SNN is implemented on an FPGA, the functionality and the performance of the chip can be tested using the testing environment shown in Fig. 4. To automate the entire testing communication between PC and FPGA, testing scripts were developed in Python using the serial interface of the computer. In addition to this, a test vector driver and a result monitor were coded in Verilog on the peripheral of the design under test, where serial data was encoded and decoded to achieve essential information. During each test sample, the test vectors are sent through UART, decoded by a driver and fed to the design as a stimulus. After one sample is finished, the final count for each class is collected and recorded on PC, where analysis and model's inference result is carried out. Two specially designed test vectors were employed, one to signal for the results to be collected and one to signal for the design to be reset. These vectors were necessary to further automate the batch emulation process.

## VI. EXPERIMENTS AND RESULTS

### A. Dataset

The dataset used in this experiment was generated from real isotope spectra obtained using CLLBC scintillation detectors. The isotopes used were $^{241}$Am, $^{133}$Ba, $^{57}$Co, $^{60}$Co, $^{137}$Cs, $^{152}$Eu, $^{226}$Ra, $^{232}$Th. The process of generating synthetic data is summarised in Fig. 5. Spectra for each isotope were collected over a range of different distances (10 *cm*, 20 *cm*, 25 *cm*, 30 *cm*, 45 *cm*, 50 *cm*, 60 *cm*, 1 *m* and 1.5 *m*). An example of the histogram with all the radioisotope under the test distance of 1 *m* is shown in Fig. 6 (left). The data recorded using different type of detectors needs to be pre-processed where all data are rebinned into data with 1024 channels. The total 105 basic isotope patterns are processed with randomly generated gain shift and additional background noise, to simulate "real world" conditions. The gain shift is mimicked by introducing a small random linear term in the calibration which is fitted with a quadratic polynomial function. Fig 6 (right) shows the synthesised Cs-137 data compared to the original. The produced synthetic dataset made of 19,005 samples provides a large number of examples with realistic variation, helping the neural network to generalise more effectively during training.

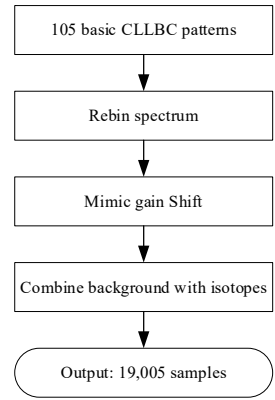

Fig. 5 Synthetic data generation flow

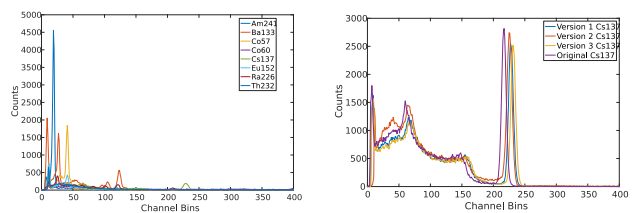

Fig. 6 (left) Data visualisation of the dataset for all 8 isotopes with integration time of 2 minutes and under the test distance of 1 *m*; (right) The original and three versions of histograms of the Cs-137 under the test distance of 25 *cm*.

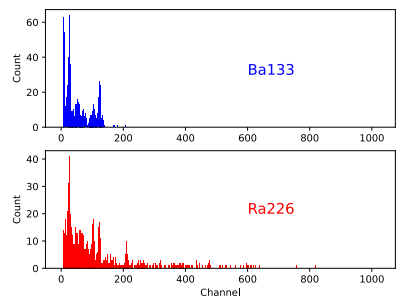

Fig. 7 Histograms of Ba-133 and Ra-226 with integration time of 3 seconds and under the test distance of 10 *cm*.

### B. CANN Results

As mentioned in section II-B, a CANN was trained via backpropagation on the histogram dataset. This training was carried out using Tensorflow [9], and the quantisation aware training available in the Tensor Model Optimization Toolkit [10]. A testing accuracy of 98.41% was achieved using an ANN model with full precision, 32-bit floating point weights which fell to 95.08% upon quantisation of weights to 8-bit signed integers. This is based on the averaged results over all five cross-validation dataset splits. Higher precision hardware would require a larger area in the FPGA implementation and so would lead to greater power consumption. For this reason we chose to trade off some accuracy for a power saving. Future work will explore this boundary between performance and power consumption.

For training of the CANNs it was necessary to normalise the data. This was done by diving the histogram counts by the sum of all channel counts, meaning that all channel counts summed to 1.

Fig. 8 shows the testing accuracy of both the full precision and quantised CANNs for parameter sweeps of both filter size and pooling size. A filter size of 5 and a pooling size of 16 were chosen because of the level of the quantised models' performance here.

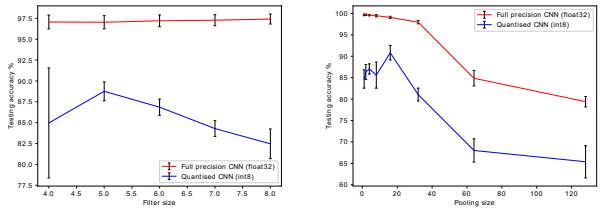

Fig. 8 Accuracy of the CANN with different filter size (left) and different pooling size (right)

### C. SpiNNaker Results

The CANN was converted to a CSNN using a version of the SNNtoolbox [6], that we adapted to be able to convert 1D convolutional and average pooling layers. The resulting CSNN model was evaluated on SpiNNaker [11] on a subset of the testing examples to validate the conversion before passing the parameters to the FPGA implementation for more rigorous evaluation.

### D. FPGA Implementation Results

The inference accuracy results of the FPGA implementation of the CSNN are shown in Fig. 9 and Fig. 10. In the 3-second testing, Ra-226 and Th-232 report very poor results. For example, Ra-226 is often identified as Ba-133 under 3-second testing due to the similarity of their histograms as illustrated in Fig. 7. This problem can be solved by increasing the testing time. As shown in Fig. 10, 132 out of 207 failed tests are passing and the accuracy is increased to 90.62% when 1 minute test time is applied.

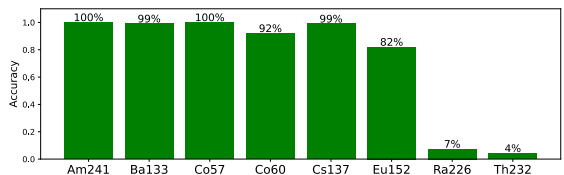

Fig. 9 Accuracies of the CSNN on FPGA with testing time of 3 seconds on 800 test dataset with 100 samples for each radioisotope.

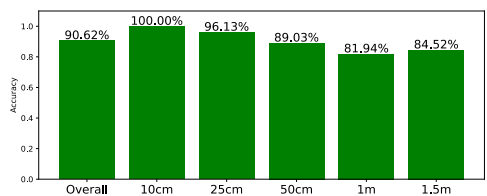

Fig. 10 Accuracies of the CSNN on FPGA with testing time of 60 seconds on 800 test dataset.

The utilization results of the implementation is shown in Table II. The SNN has been implemented on the Xilinx Artix-7 FPGA chip and the system is running with a clock frequency of 100 MHz.

The average power consumption per inference is about 75 $mW$. Xilinx Vivado power analysis tool is used to estimate the power consumption. The Switching Activity Interchange format (SAIF) file, which is produced from the post-implementation functional simulations, is applied to enhance the power estimation accuracy. Table III shows the breakdown of the power consumption. The major power cost is from static power and clocks. The processing power cost is less than 2% of total consumption.

TABLE II. LOW-LEVEL DEVICE UTILIZATION OF THE SNN

| Logic Utilisation | Used | Available | Utilisation |
| --- | --- | --- | --- |
| Slice Registers | 1,470 | 41,600 | 3.53% |
| Slice LUTs | 1,044 | 20,800 | 5.02% |
| BUFGCTRL | 1 | 32 | 3.13% |
| Block RAM | 1 | 50 | 2% |

TABLE III. BREAKDOWN OF ON-CHIP POWER CONSUMPTION

| Dynamic Power ($mW$) | | | | | Static Power ($mW$) |
| --- | --- | --- | --- | --- | --- |
| Clocks | Signals | Logic | BRAM | I/O | |
| 4 | < 1 | < 1 | < 1 | < 1 | 70 |

### E. Accuracy Comparisons

To compare with other machine learning algorithms, three other algorithms were implemented on the same data set. The comparison results are shown in table IV. The full precision CANN outperforms other machine learning, and less than 3% accuracy loss can be traded off for quantisation with more power and area saving. There is less than 5% accuracy drop from CANN to CSNN. However, 74% of the tests are passing within 3 second and with half of the test time, the CSNN achieves accuracy of more than 90%.

TABLE IV. ACCURACY COMPARISON

| Algorithms | Platform | Accuracy | Integration Time |
| --- | --- | --- | --- |
| $k$-Nearest Neighbour | CPU | 97.24% | 120 s |
| Decision Tree | CPU | 92.14% | 120 s |
| Random Forest | CPU | 97.67% | 120 s |
| CANN (float32) | GPU | 98.41% | 120 s |
| CANN (int8) | GPU | 95.08% | 120 s |
| CSNN | FPGA | 90.62% | < 60 s |

## VII. CONCLUSION

An event-based hardware design for low-power and high resolution radioisotope identification has been presented. This paper has detailed the methodology from initial design to FPGA implementation and presented results. For ease of comparison between different isotopes and distances, normalised histograms were used as inputs to models here. Future works will expand the techniques developed here by looking at the problem of multi-label classification of mixed radioisotopes, exploring asynchronous methods to further reduce power and will look at the implementation of plasticity and learning within the hardware implementation of SNNs, using unsupervised Spiking-Timing-Dependent Plasticity (STDP) to enhance the ability of the system to deal with new data.

ACKNOWLEDGMENT

This project is funded by US Defense Threat Reduction Agency (DTRA) and Kromek Group plc, UK.